\documentclass[preprint,amsmath,amssymb]{revtex4}
\usepackage{graphicx,dcolumn,bm}
\usepackage[bf,SL,BF]{subfigure}
\usepackage{txfonts}

\begin{document}

\title{Nonlinear thermal properties of three-terminal mesoscopic dielectric systems}

\author{Y. Ming}
\email{meanyee@ustc.edu}
 \affiliation{School of Physics and Material Science, Anhui
 University,\\
 Hefei, Anhui 230039, People's Republic of China}

\author{Z.X. Wang}
\author{Q. Li}
\author{Z.J. Ding}
\email{zjding@ustc.edu.cn}
\affiliation{
Hefei National Laboratory for Physical Sciences at Microscale and\\
Department of Physics, University of Science and Technology of
China,\\
Hefei, Anhui 230026, People's Republic of China
}%

\date{\today}

\begin{abstract}
This paper studies the thermal properties of three-terminal
mesoscopic dielectric systems in the nonlinear response regime at
low temperature. For a symmetric three-terminal system, when the
temperature is finitely different between the left and right thermal
reservoir, the temperature of the central thermal reservoir is
always higher than the averaging temperature of the others. This
nonlinear thermal phenomenon is also observed for asymmetric
three-terminal systems. At the end, a model of thermal rectification
is presented.
\end{abstract}

\maketitle

With the development of the modern electronics, heat conduction, as
the counterpart of electric conduction, has attracted much attention
in recent years. Some research works
\cite{Liwx04,Liwx06,Tang06,Peng07} focus on the universal quantum of
thermal conductance which is predicted by Rego and Kirczenow
\cite{Rego98} and has already been verified by experiment
\cite{Schwab00}. The quantum of thermal conductance indicates that
the heat conduction is determined by the ballistic transmission of
the acoustic phonon at low temperature. On the other hand, thermal
rectification to control the heat flux
\cite{Terraneo02,Li04,Li05,Li06,Hu06,Segal05,Eckmann06, Casati07} is
very interesting. Experimental work to demonstrate the thermal
rectification is reported recently \cite{Chang06}.

Based on the above efforts, this paper further studies thermal
rectification in ballistic heat conduction. For heat conduction in a
two-terminal mesoscopic dielectric system at low temperature,
because transmission coefficients of ballistic phonons are
independent on the temperature, there should be no thermal
rectification\cite{Segal05}. But, for electric conduction within a
three-terminal ballistic junction, previous investigations have
indicated the nonlinear ballistic transport of electrons. That is,
if voltages $V_L$ and $V_R$ are applied on the left and right
branches of a symmetric three-terminal ballistic junction in
push-pull fashion, with $V_L=-V_R$, the voltage at the central
branch is always Negative
\cite{Hieke00,Xu01,Shorubalko01,Worschech01,Csontos02,Csontos03,
Wallin06}. This nonlinear property can be used for rectification,
second-harmonic generation, and logic function
\cite{Xu02,Shorubalko03,Worschech05}. Motivated by the nonlinear
electrical properties in three-terminal ballistic junctions, in this
work, we study the thermal properties of three-terminal mesoscopic
dielectric systems in the nonlinear response regime and try to
propose a model of thermal rectification. The model works at low
temperature in order to keep the ballistic transmission of phonons,
similar with the model taken in Ref.~\cite{Casati07}.

\begin{center}
\begin{figure}[htbp]
\includegraphics[width=0.3\textwidth]{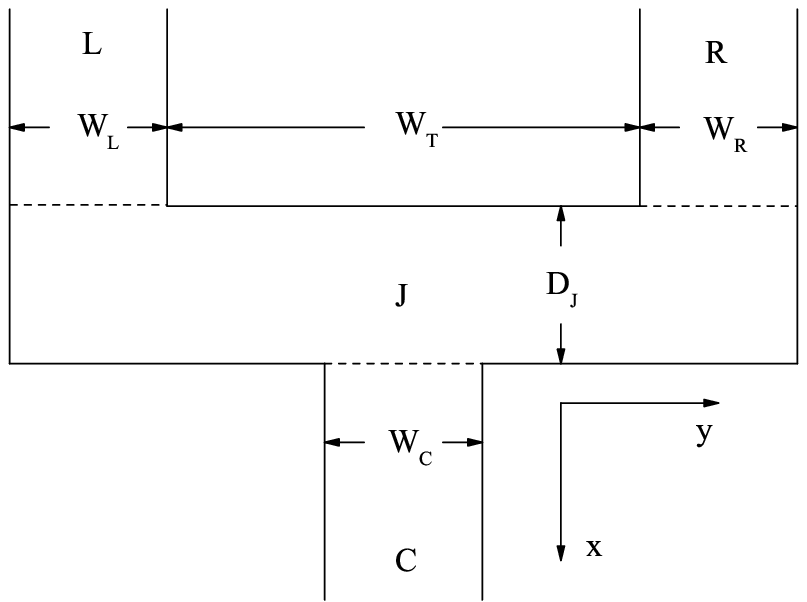}
\caption{\label{fig1}}
\end{figure}
\end{center}

The geometry of the symmetric three-terminal mesoscopic dielectric
system is sketched in Fig.~\ref{fig1}. Regions -$L$, -$R$ and -$C$
are left, right and central terminals, respectively. Region -$J$ is
the midsection. Assuming that the terminals are perfect and phonons
coming from thermal reservoirs are not scattered within the
terminals, the energy flux $\dot{Q}_i$ from terminal $i$ ($i=L,R,C$)
flowing into the midsection $J$ can be expressed as
\cite{Sun02,Yang04}
\begin{equation}\label{eqn1}
    \dot{Q}_i=\sum_{j (j\neq i)} \sum_m \int_{\omega_{im}}^{+\infty}
    [n(T_i,\omega)-n(T_j,\omega)]\hbar\omega\, \tau_{ji,m}(\omega) \frac{d\omega}{2\pi},
\end{equation}
where $n(T_i,\omega)=[\exp (\hbar \omega/k_B T_i)-1]^{-1}$ is the
Bose-Einstein distribution function of the phonons in the $i$th
reservoir, $T_i$ is the equilibrium temperature of thermal reservoir
$i$, $\omega_{im}$ is the cutoff frequency of mode $m$ in terminal
$i$, $\tau_{ji,m}(\omega)=\sum_n \theta (\omega -\omega_{jn})
\tau_{ji,nm}$ and $\tau_{ji,nm}$ is the transmission coefficient
from mode $m$ of terminal $i$ at frequency $\omega$ across all the
interface into the mode $n$ of terminal $j$.

Firstly we let the energy flux flows through the two terminals $L$
and $R$ and keep zero flux in terminal $C$ (i.e. $\dot{Q}_C=0$) and
then calculate the temperature $T_C$. In the linear response regime,
the flux in terminal $C$ can be written as \cite{Sun02, Yang04}
$\dot{Q}_C=\sum_{j=L,R} G_{jC}\, (T_C-T_j)=0$, where $G_{jC}$ is the
two-terminal thermal conductance from central terminal $C$ to
terminal $j$ ($=L$ or $R$). Due to the symmetry of the
three-terminal mesoscopic dielectric system and the independence on
the temperature of transmission coefficients, the thermal
conductances from central terminal to left terminal and right
terminal are equal, i.e. $G_{LC}=G_{RC}$. So the temperature of
central thermal reservoir in the linear response regime is simply
the averaging temperature of the left and the right thermal
reservoirs, i.e. $T_C=(T_L+T_R)/2$.

When the system is not in the linear response regime and with the
finite temperature difference of $2|\Delta T|$ between the left
thermal reservoir and the right thermal reservoir, what is the
temperature $T_C$ of the central thermal reservoir? To figure out
it, we let $T_L=T_0+\Delta T$ and $T_R=T_0-\Delta T$. By using the
Taylor expansion of the Bose-Einstein distribution function
$n(T_i,\omega)$, when $|\Delta T|$ is a small value, the temperature
$T_C$ can be written as
\begin{eqnarray}\label{eqn2}
T_C=T_0+\frac{1}{2}\alpha (\Delta T)^2+\mathcal {O} [(\Delta T)^4],
\end{eqnarray}
where
\begin{eqnarray}\label{eqn3}
\alpha=\frac{\sum_m \int_{\omega_{Cm}}^{+\infty}\left(
\frac{\partial^2 n(T,\omega)} {\partial T^2} \right)_{T_0}\hbar
\omega \tau_{LC,m}(\omega) \frac{d\omega}{2\pi}} {\sum_m
\int_{\omega_{Cm}}^{+\infty}\left( \frac{\partial n (T,\omega)}
{\partial T} \right)_{T_0}\hbar \omega \tau_{LC,m}(\omega)
\frac{d\omega}{2\pi}}.
\end{eqnarray}
Here, $\tau_{RC,m}= \tau_{LC,m}$ by the symmetry of the system and
the independence on the temperature of the transmission
coefficients. Thus, $T_C$ depends quadratically on $\Delta T$ for
small $|\Delta T|$. By using $e^x=\sum_{l=0}^\infty
\frac{1}{l!}x^l$, for all $\omega>0$, we can easily obtain that
\begin{eqnarray}
\left(\frac{\partial n(T,\omega)}{\partial T}\right)_{T_0} &=& \frac
{\hbar \omega \exp (\hbar \omega/k_B T_0)} {k_B T_0^2 [\exp (\hbar
\omega/k_B T_0)-1 ]^2}>0,\label{eqn4}\\
\left(\frac{\partial^2 n(T,\omega)}{\partial T^2}\right)_{T_0} &=&
\frac {\hbar \omega \exp (\hbar \omega/k_B T_0)} {k_B T_0^3 [\exp
(\hbar \omega/k_B T_0)-1 ]^3}\nonumber \\ && \times
\sum_{l=3}^\infty \frac{l-2} {l!} \left(\frac{\hbar\omega}{k_B
T_0}\right)^l>0.\label{eqn5}
\end{eqnarray}
Thus, it is obtained that $\alpha>0$ and $\alpha\propto 1/T_0$
approximately. Thus $T_C>T_0=(T_L+T_R)/2$ is always true when the
temperature difference between the left and the right thermal
reservoirs is finite.

Next, we will carry out numerical calculations for a three-terminal
system shown in Fig.~\ref{fig1}. The scalar model for the elastic
wave is considered. And the model for thin geometry at low
temperature is used so that the calculation is two-dimensional. So
we can derive the transmission coefficient, $\tau_{ji,m}$, by the
scattering matrix method \cite{Sun02,Sheng97,Ming06}. In the
calculation, we employ the following values of elastic stiffness
constant and the mass density for GaAs\cite{Chen00}:
$C_{44}=5.99\times 10^{10}$ Nm$^{-2}$ and $\rho=5317.6$ kgm$^{-3}$,
and choose $W_L=W_R=10$~nm, $W_T=20$~nm and $W_C=D_J=10$~nm. We
truncate the sum of $m$ in Eq.~(\ref{eqn1}) at $m=10$ \cite{Ming06}
and limit the temperatures of the left and the right thermal
reservoirs lower than $T_{ph} =\hbar \pi v/W_L k_B \approx 7.61$~K
($v=\sqrt{C_{44}/\rho}$ is the sound velocity). At this low
temperature, the phonon relaxation can be neglected \cite{Sun02} and
the heat conduction is mainly determined by the ballistic
transmission of the acoustic phonons.

\begin{center}
\begin{figure}[htbp]
\includegraphics[width=0.5\textwidth]{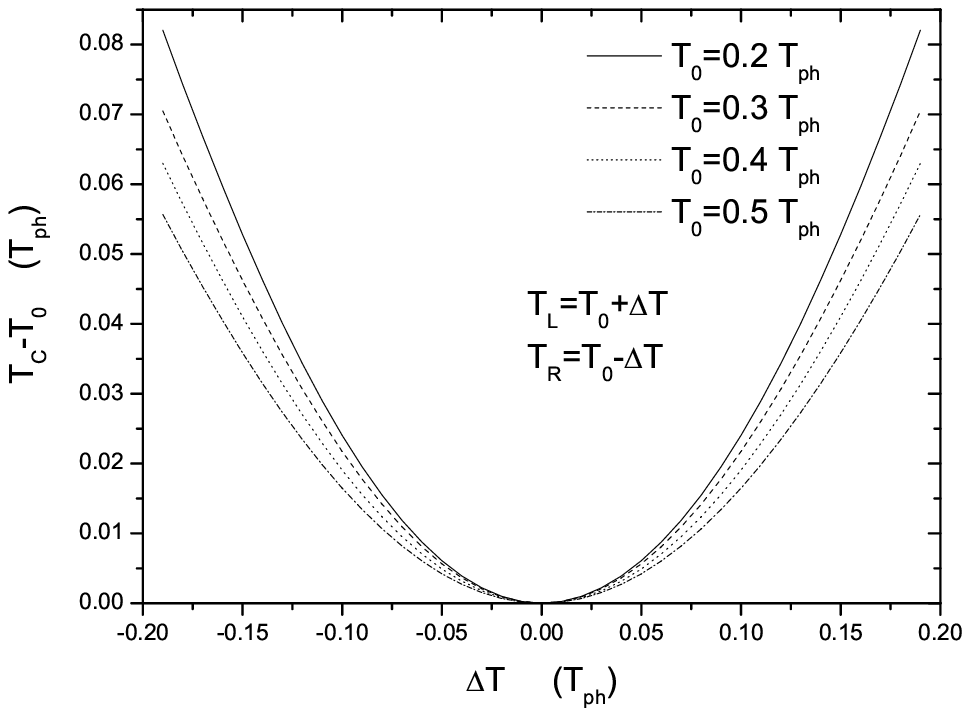}
\caption{\label{fig2}}
\end{figure}
\end{center}

Fig.~\ref{fig2} shows $T_C-T_0$ vs $\Delta T$ for four different
$T_0$s, where $T_0$ is the averaging value of $T_L$ and $T_R$, and
$T_L=T_0+\Delta T$, $T_R=T_0-\Delta T$. First, it can be seen that
when the difference between $T_L$ and $T_R$ is finite ( $|2\Delta
T|>0$), the temperature of the central thermal reservoir $T_C$ is
always higher than the averaging value of $T_L$ and $T_R$, i.e.
$T_C-T_0>0$, no matter which thermal reservoir has the higher
temperature. Second, the temperature $T_C$ shows a quadratic
dependence on $\Delta T$, in agreement with Eq.~(\ref{eqn2}). Third,
as mentioned above that $\alpha \propto 1/T_0$ approximately, the
curvatures of the curves depend strongly on the temperature $T_0$.
The lower the temperature $T_0$, the larger the curvature.

To study the nonlinear properties in the asymmetric three-terminal
systems, we let the energy flux flows through the two terminals $L$
and $C$ and keep zero flux in terminal $R$ (i.e. $\dot{Q}_R=0$) and
then calculate the temperature $T_R$. Same as Eq.~(\ref{eqn2}), with
$\tau_{CR,m}>\tau_{LR,m}$, the temperature $T_R$ can be written as
\begin{eqnarray}\label{eqn6}
T_R=T_0-\beta (\Delta T)+\frac{1}{2}\gamma (\Delta T)^2+\mathcal {O}
[(\Delta T)^3],
\end{eqnarray}
where
\begin{eqnarray}
\beta=\frac{\sum_m \int_{\omega_{Rm}}^{+\infty}\left( \frac{\partial
n(T,\omega)} {\partial T} \right)_{T_0}\hbar \omega
(\tau_{CR,m}-\tau_{LR,m}) \frac{d\omega}{2\pi}} {\sum_m
\int_{\omega_{Rm}}^{+\infty}\left( \frac{\partial n (T,\omega)}
{\partial T} \right)_{T_0}\hbar \omega (\tau_{CR,m}+\tau_{LR,m})
\frac{d\omega}{2\pi}}>0,\label{eqn7}\\
\gamma=\frac{\sum_m
\int_{\omega_{Rm}}^{+\infty}\left( \frac{\partial^2 n(T,\omega)}
{\partial T^2} \right)_{T_0}\hbar \omega (\tau_{CR,m}+\tau_{LR,m})
\frac{d\omega}{2\pi}} {\sum_m \int_{\omega_{Rm}}^{+\infty}\left(
\frac{\partial n (T,\omega)} {\partial T} \right)_{T_0}\hbar \omega
(\tau_{CR,m}+\tau_{LR,m}) \frac{d\omega}{2\pi}}>0.\label{eqn8}
\end{eqnarray}

\begin{center}
\begin{figure}[htbp]
\includegraphics[width=0.5\textwidth]{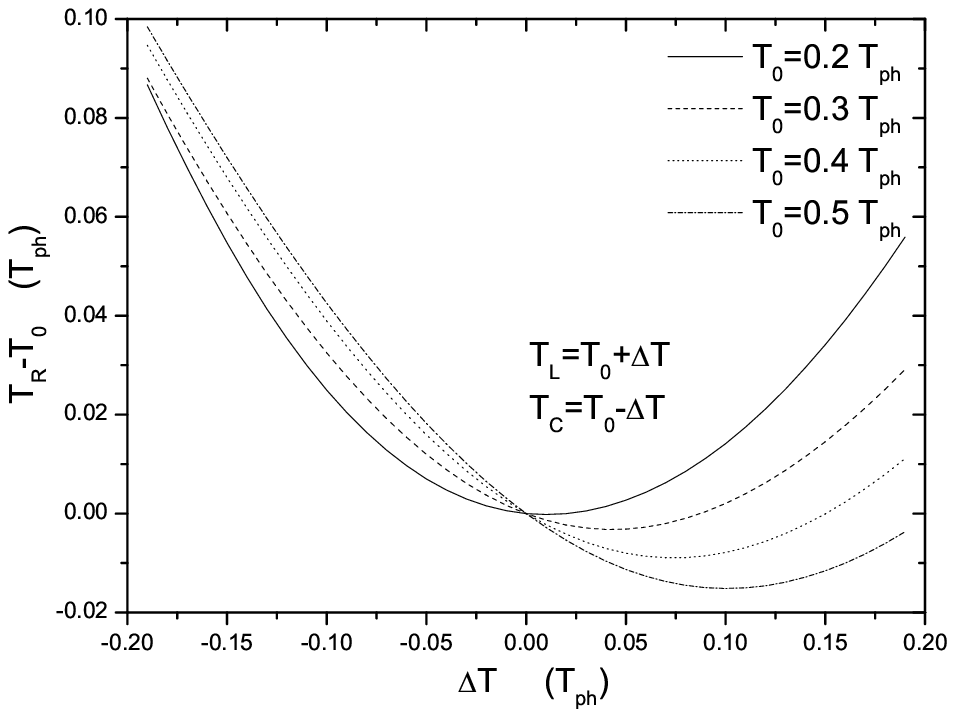}
\caption{\label{fig3}}
\end{figure}
\end{center}

Fig.~\ref{fig3} shows $T_R-T_0$ vs $\Delta T$ for four different
$T_0$s. The curves are also open up parabolic, in agreement with
Eq.~(\ref{eqn6}). The major difference with the symmetric case is
that, here $T_R-T_0$ is no longer a symmetric function of $\Delta T$
with respect $\Delta T=0$ but respect about $\Delta T=\beta/\gamma$,
and can be less than zero when $0<\Delta T<2\beta/\gamma$. This
means for some specific positive values of $\Delta T$, the
temperature $T_R$ can be less than the averaging value of $T_L$ and
$T_C$. The minimal value of $T_R$ is about $T_0-\beta^2/2\gamma$.

In summary, for symmetric three-terminal systems in the nonlinear
response regime, we have found that the temperature of the central
thermal reservoir is always higher than the averaging temperature of
the left and the right thermal reservoirs. For the asymmetric
three-terminal systems, the same nonlinear thermal properties can be
also observed except that, in some special situations, the
temperature of the central thermal reservoir can be lower than the
averaging temperature of the left and the right thermal reservoirs.
We would like emphasize that the temperature of the central thermal
reservoir is insensitive to the details of the transmission
characteristics. It is different from the electric case where the
output voltage at the central branch shows fluctuations due to the
transmission fluctuations\cite{Csontos03}. This difference is due to
the fact that the electric current is carried by a few electrons
near the Fermi energy\cite{Datta97} but the heat flux is contributed
by all phonons with different frequencies. The nonlinear thermal
properties of the symmetric three-terminal systems can be used to
control the heat flux. For example, when the temperature $T_C$ of
the central thermal reservoir is lower than the averaging
temperature $(T_L+T_R)/2$ of the left and the right thermal
reservoirs, it is easier that the energy flux flows into the central
thermal reservoir than the energy flux flows out from the central
thermal reservoir when $T_C>(T_L+T_R)/2$.

Y. Ming is supported by the Talent Group Construction Foundation of
Anhui University (Grant No.~02203104/04). Z.J. Ding is supported by
the National Natural Science Foundation of China (Grant No. 10574121
and 10025420), Chinese Education Ministry and Chinese Academy of
Sciences.

\newpage

Figure Captions:

Fig.~\ref{fig1}. Schematic illustration of a symmetric
three-terminal mesoscopic dielectric system.

Fig.~\ref{fig2}. $T_C-T_0$ vs $\Delta T$, calculated for a
three-terminal system shown in Fig.~\ref{fig1}, for four different
averaging temperatures, $T_0=(T_L+T_R)/2$, where $T_C$, $T_L$ and
$T_R$ are the temperatures of the central, the left and the right
thermal reservoirs, respectively. All the temperatures are reduced
by $T_{ph}\approx 7.61$~K.

Fig.~\ref{fig3}. $T_R-T_0$ vs $\Delta T$, calculated for a
three-terminal system shown in Fig.~\ref{fig1}, for four different
temperatures, $T_0=(T_L+T_C)/2$, where $T_C$, $T_L$ and $T_R$ are
the temperatures of the central, the left and the right thermal
reservoirs, respectively. All the temperatures are reduced by
$T_{ph}\approx 7.61$~K.

\end{document}